\documentclass{article}
\usepackage{bbding}
\usepackage{spconf,amsmath,graphicx,hyperref}
\usepackage{siunitx}
\usepackage{booktabs}
\usepackage{multirow} 
\usepackage{amssymb}
\usepackage{cite}
\usepackage{amsfonts}
\usepackage{array}
\usepackage[caption=false,font=normalsize,labelfont=sf,textfont=sf]{subfig}
\usepackage{textcomp}
\usepackage{makecell}
\usepackage{url}
\usepackage{verbatim}
\usepackage{enumitem}
\usepackage{makecell}
\usepackage{etoolbox}
\robustify\bfseries
\usepackage{tabularx}
\usepackage{color}
\usepackage{arydshln} 
\def\H{{\mathsf H}}

\usepackage{caption}
\usepackage{tikz}

\makeatletter
\renewcommand\normalsize{%
   \@setfontsize\normalsize{9pt}{10.5pt}} 
\makeatother
\let\OLDthebibliography\thebibliography
\renewcommand\thebibliography[1]{
  \OLDthebibliography{#1}
  \setlength{\parskip}{0.5pt}
  \setlength{\itemsep}{1pt plus 0.3ex}
}
\usepackage{flushend}

\title{
UltraM2M: Leveraging Text Transcripts and Mixture Constraints for Weakly-Supervised Speech Enhancement
}
\name{Jiachen Liu, Fulin Wu, and Zhong-Qiu Wang
}
\address{
Southern University of Science and Technology, Shenzhen, China\\
{\small\texttt{\{liu.ameth, wang.zhongqiu41\}@gmail.com}
}
}
\begin{document}
%
\maketitle
\begin{abstract}
We propose UltraM2M, a weakly-supervised speech enhancement algorithm building upon the recent unsupervised mixture-to-mixture (M2M) algorithm.
M2M realizes unsupervised speech enhancement by training deep neural networks on a set of real-recorded noisy-reverberant multi-channel mixture signals to estimate target speech and non-target signals. The two estimated signals are penalized by a so-called mixture-constraint (MC) loss, which constrains them to reconstruct the observed mixture signals.
Although shown to be effective, the mixture constraint may be too weak to enable sufficient noise reduction.
To deal with this, UltraM2M extends M2M by further leveraging text transcripts of real-recorded mixtures to design an automatic speech recognition (ASR) loss to penalize the estimated speech signal.
The ASR loss can be viewed as a form of weak supervision that could help unsupervised enhancement.
Evaluation results on the CHiME-$4$ dataset show the effectiveness of UltraM2M.
\end{abstract}
\section{Introduction}\label{sec:intro}

The current dominant approach for speech enhancement (SE) is supervised learning, where clean speech signals are synthetically mixed with noise and reverberation to create simulated pairs of clean speech and noisy-reverberant mixture signals for training supervised models such as deep neural networks (DNNs) \cite{huang2025advances, yurtay2026comprehensive}.
Although obtaining strong SE performance on simulated test data whose distribution is similar to that of the simulated training set, the trained models often exhibit limited generalizability to real-recorded mixture signals\cite{nugraha2016analysis, zhang2021closing, Zhang2022monoaural, cui2025reducing}, largely due to the mismatches between simulated training and real-recorded test conditions.
In this context, unsupervised or weakly-supervised SE, which trains DNN models directly on real-recorded data in the target domain, has shown strong potential in addressing such mismatches \cite{saleem2020improvement, ochieng2023deep, leglaive2023udase, nortier2024unsupervised, pishdadian2020finding, maciejewski2026exploiting, Chang2019MIMOSPEECH,xu2025adps}.
Compared with unsupervised methods, weakly-supervised SE can obtain better performance, as it has some supervision to leverage during training \cite{leglaive2023udase, nortier2024unsupervised, maciejewski2026exploiting, pishdadian2020finding, Chang2019MIMOSPEECH,xu2025adps}.

Among various types of weakly-supervised SE methods \cite{maciejewski2026exploiting, pishdadian2020finding, Chang2019MIMOSPEECH,xu2025adps}, using automatic speech recognition (ASR) models to help train SE models on real-recorded data is a practical choice \cite{Chang2019MIMOSPEECH}, since annotating real-recorded mixtures to produce the text transcripts required by ASR models is inexpensive and straightforward.
ASR systems, however, are complicated: their effectiveness in improving SE is highly sensitive to how they are exploited during training, and different strategies can yield remarkably different enhancement performance.
Therefore, steering ASR systems toward an operating regime that reliably improves SE remains a non-trivial challenge.

To exploit the weak supervision afforded by text transcripts, a natural strategy is \textit{end-to-end joint training}.
Some studies first pre-train an SE model and then jointly train it with an ASR model, optimizing only the ASR loss \cite{Wang2015aJointTraining, Wang2016TASLP, masuyama2025csl}.
Although better ASR performance is reported, a practical limitation is that the ASR models used at run time are often owned by third parties and closed-source, and therefore cannot be used to train SE models.
In addition, end-to-end training causes the ASR model to adapt to artifacts introduced by the SE model, making the two tightly coupled \cite{iwamoto2022bad,iwamoto2024e2e}.
This often renders the enhanced speech less useful for other downstream tasks.
Another strategy is \textit{multi-task training}, which optimizes both the enhancement and the ASR loss.
There are various ways to implement it, such as alternative training \cite{eskimez2021human}, gradient remedy training \cite{hu2023gradient, lee2023d4am}, and feature fusion training \cite{fan2020gated}.
In these studies, the SE branch is usually designed for supervised training and thus assumes the availability of clean target speech.
Such supervision, however, is unavailable for real-recorded mixtures, which typically lack paired clean speech.

M2M training \cite{wang2024m2m} trains DNN to produce an estimate for target speech and another for non-target signals, such that the two estimates, after being properly linearly-filtered and summated, can reconstruct the observed mixtures.
Building on this scheme, we propose UltraM2M, which extends M2M by further exploiting text transcripts.
Specifically, UltraM2M employs pretrained, frozen ASR models to exploit the text transcripts of real-recorded data.
In addition to training on real-recorded data, UltraM2M also inherits the co-learning strategy from SuperM2M \cite{wang2025superm2m}, which trains the same DNN model by alternating between simulated and real-recorded mixtures.
Unlike SuperM2M, where the supervised training aims to avoid source permutation, frequency permutation, and source ambiguity, the supervised training in UltraM2M serves more as an auxiliary constraint.
We therefore pack each mini-batch with both simulated and real-recorded mixtures, so that both data types contribute to the same parameter update.
UltraM2M thus allows us to exploit the weak supervision from text transcripts, which can be obtained through inexpensive annotation, while mitigating its potential drift by regularizing it with the MC loss and with supervised training on simulated mixtures.
On the CHiME-$4$ dataset \cite{barker2016chime4}, UltraM2M obtains much better WER than SuperM2M on the real-recorded sets, as well as much better speech quality in multi-channel settings.

\section{Review of M2M}\label{M2M_description}

\begin{figure*}
    \centering
    \includegraphics[width=\linewidth]{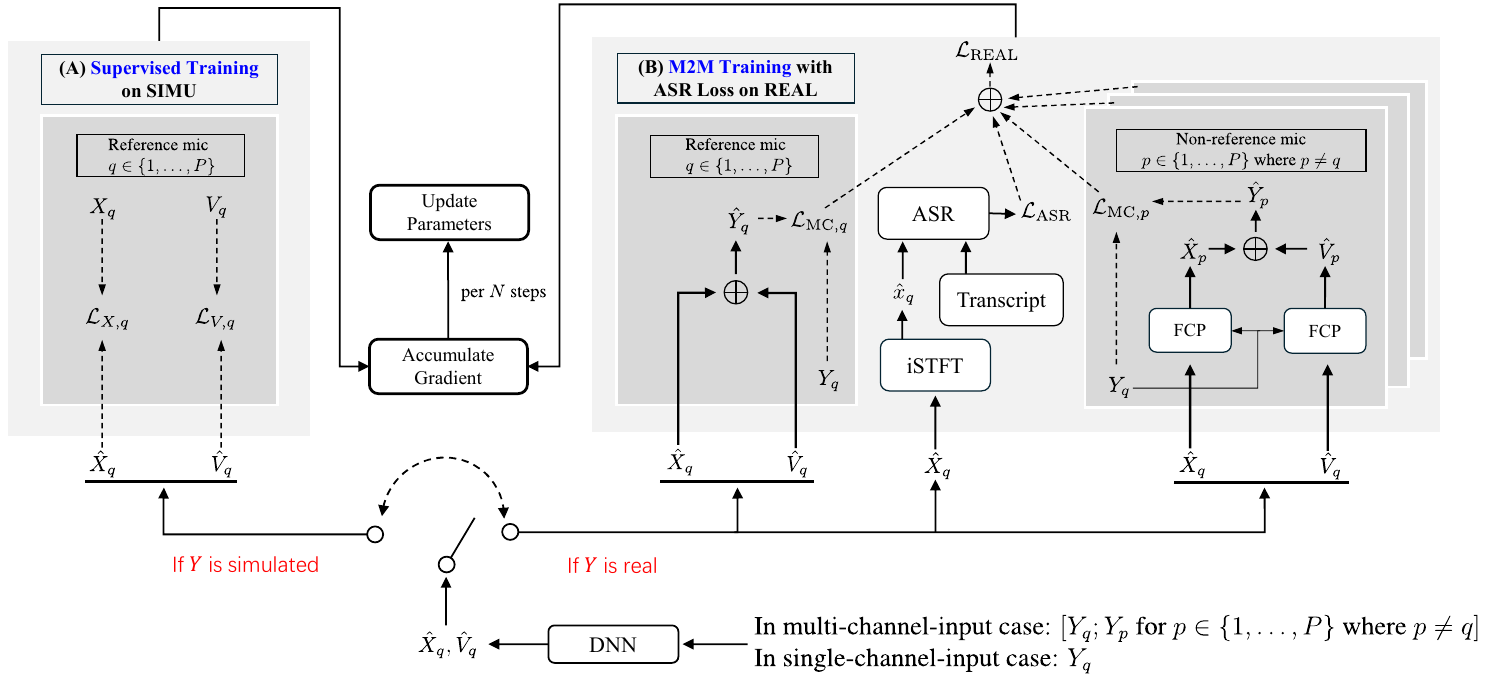}
    \caption{Overview of UltraM2M training. (A) supervised training on simulated mixtures; and (B) M2M training with ASR loss on real-recorded mixtures.
    }
    \label{fig:ultram2m_framework}
\end{figure*}

This section reviews M2M \cite{wang2024m2m}.
In a noisy-reverberant environment with a compact $P$-microphone far-field array and a single target speaker, the physical model for each recorded mixture can be formulated, in the short-time Fourier transform (STFT) domain, as $Y_p(t,f) = X_p(t,f) + V_p(t,f)$,
where $Y_p(t,f)$, $X_p(t,f)$, and $V_p(t,f)$ respectively denote the STFT coefficients of the mixture, reverberant speech of the target speaker, and non-speech signals captured by the far-field microphone $p$ at time $t$ and frequency bin $f$.
We aim to suppress non-speech signals and reconstruct the reverberant target speech at a reference microphone $q$ (i.e., $X_q$), based on the multi-channel input mixture $\mathbf{Y} = \{Y_p\}_{p=1}^P$.

M2M is designed to train unsupervised SE models solely based on a set of real-recorded mixtures.
To employ M2M training, a DNN takes in far-field mixtures as input and produces $\hat{X}_q$ and $\hat{V}_q$ respectively for the target signal and non-target signals at the reference microphone $q$.
Each estimated signal is then linearly filtered by forward convolutive prediction (FCP) \cite{wang2021convolutive} and used to optimize a mixture-constraint (MC) loss, which enforces the two DNN-estimated signals to be able to reconstruct each of the far-field mixture signals.
The MC loss can be defined as follows:
\begin{equation}
    \mathcal{L}_{\text{MC}} = \mathcal{L}_{\text{MC},q} + \frac{1}{P-1}  \sum\nolimits_{p=1, p\neq q}^{P}  \mathcal{L}_{\text{MC},p},
    \label{eq:mixture_constraint_loss}
\end{equation}
where $\mathcal{L}_{\text{MC},q}$ and $\mathcal{L}_{\text{MC},p}$ are respectively the MC losses at the reference microphone and each non-reference microphone.
At the reference microphone $q$, the sum of the two estimates is constrained to be equal to the mixture:
\begin{equation}
    \mathcal{L}_{\text{MC},q} = \sum\nolimits_{t,f} \mathcal{F}\Big(Y_q(t,f),\; \hat{X}_q(t,f) + \hat{V}_q(t,f)\Big),
\end{equation}
where the distance function $\mathcal{F}$ is defined as
\begin{align} 
    \mathcal{F}(a,b) &= {\mathcal{G}(a,b)} \big/ \sum\nolimits_{t',f'}|a(t',f')|, \\
    \mathcal{G}(a,b) &= \big|\mathcal{R}(a)-\mathcal{R}(b)\big| + \big|\mathcal{I}(a)-\mathcal{I}(b)\big| + \big||a|-|b|\big|, \label{eq:gfunction}
\end{align}
with $\mathcal{R}(\cdot)$, $\mathcal{I}(\cdot)$, and $|\cdot|$ respectively extracting the real, imaginary, and magnitude components.
At each non-reference microphone $p$ ($\neq q$), since the estimates are produced at microphone $q$, multi-frame linear filtering based on FCP \cite{wang2021convolutive} is applied to account for inter-channel differences:
\begin{equation}
\mathcal{L}_{\text{MC},p} =\sum\nolimits_{t,f} \mathcal{F}\Big( Y_p(t,f), 
\hat{\mathbf{h}}_p(f)^\H \overline{\hat{\mathbf{X}}}_q(t,f) + \hat{\mathbf{r}}_p(f)^\H \overline{\hat{\mathbf{V}}}_q(t,f)
\Big), \label{MC_loss_nonref}
\end{equation}
where ${\overline{\hat{\mathbf{X}}}}_q(t,f) = [\hat{X}_q(t-I+1,f),\dots,\hat{X}_q(t+J,f)]^{\mathsf{T}} \in \mathbb{C}^{I + J}$ stacks a window of $I+J$ T-F units of $\hat{X}_q$, ${\overline{\hat{\mathbf{V}}}}_q(t,f)$ is analogously defined for $\hat{V}_q$, and the filter taps $I$ and $J$ are hyper-parameters.
Taking $\hat{\mathbf{h}}_p (f)$ as an example, the filter can be computed by FCP \cite{wang2021convolutive} by solving the following minimization problem:
\begin{align}
    \hat{\mathbf{h}}_p (f) = \underset{\textbf{h}_p(f)}{\text{argmin}} \sum\limits_{t} 
    \frac{|Y_p(t,f) - \textbf{h}_p(f)^\H \overline{\hat{\textbf{X}}}_q (t,f)|^2}{\hat{\lambda}_p(t,f)},
\end{align}
where $\hat{\lambda}_p(t,f) = \xi \cdot \max(|Y_p|^2) + |Y_p(t,f)|^2$ floors the denominator, and $\xi$ is a hyper-parameter to tune.

\section{UltraM2M}
\label{ultram2m}

This section presents UltraM2M, which (a) trains the DNN on real-recorded mixtures with a frozen ASR loss and the MC loss, (b) keeps a supervised loss on simulated mixtures, and (c) integrates the two branches by using both simulated and real-recorded mixtures for model training.
See Fig. \ref{fig:ultram2m_framework} for an illustration.

\subsection{Weakly-Supervised Training on Real-Recorded Mixtures}\label{ultram2m_weakly}
Fig. \ref{fig:ultram2m_framework} (B) illustrates UltraM2M weakly-supervised training on real-recorded data.
For each real mixture, the enhancement model takes a fixed number of microphones of the input mixture as its input and produces a target signal estimate and a non-target signal estimate.
Based on the two estimates, we take all microphones and use them to compute the MC loss $\mathcal{L}_{\text{MC}}$ defined in Section \ref{M2M_description}.  
Note that the MC loss constrains only the sum of the two estimates, and thus cannot by itself tell which estimate corresponds to the target speech; it therefore provides only a weak constraint for speech enhancement.
Using a pretrained, frozen ASR model, an ASR loss $\mathcal{L}_{\text{ASR}}$ is employed to resolve this ambiguity by penalizing the designated speech estimate when it does not match the annotated transcript.
Here we designate the first output as the speech estimate, which is consistent with the following supervised training branch.
UltraM2M is not tied to a particular ASR model or ASR loss, so it can be instantiated with any pretrained ASR backend.
The loss for training on real-recorded mixtures is defined as
\begin{equation}
    \mathcal{L}_{\text{REAL}} = \mathcal{L}_{\text{MC}} + \beta \times \mathcal{L}_{\text{ASR}},
    \label{eq:real_loss}
\end{equation}
where $\mathcal{L}_{\text{MC}}$ is defined in Eq. \eqref{eq:mixture_constraint_loss} and $\beta$ controls the weight of the ASR loss.

The MC loss is computed from all the microphones of the training mixture, whereas the DNN can take only a subset of the microphones as its input.
For example, we can feed the mixture at the reference microphone to the DNN, while computing the MC loss over all the microphones.
Moreover, the ASR loss also only requires the DNN output and transcripts.
This way, the DNN can perform monaural enhancement at run time.

\subsection{Supervised Training on Simulated Mixtures}\label{ultram2m_supervised}

For simulated mixtures, the DNN is trained with a supervised loss as an auxiliary constraint:
\begin{align}
    \mathcal{L}_{X,q} &= 
    \frac{1}{{\sum_{t,f}|Y_q(t,f)|}}{\sum\nolimits_{t,f} \mathcal{G}\big(X_q(t,f), \hat{X}_q(t,f)\big)}, \\
   \mathcal{L}_{V,q} &=  \frac{1}{\sum\nolimits_{t,f}|Y_q(t,f)|}{\sum\nolimits_{t,f} \mathcal{G}\big(V_q(t,f), \hat{V}_q(t,f)\big)},
\end{align}
where $X_q$ and $V_q$ are respectively the ground-truth reverberant speech and non-speech components of the simulated mixture, and the normalization and the distance $\mathcal{G}(\cdot)$ are the same as in Eq.~\eqref{eq:gfunction}.

\subsection{
Batch Training with Both Simulated and Real Mixtures 
}\label{sec:accumulation_setup}

During training, we compute each parameter update using both simulated and real-recorded mixtures, rather than using only one type of mixture per training step and alternating between the two losses across steps \cite{eskimez2021human}.
Under an alternating schedule, an update is constrained by only one of the two objectives; mixing both types in every update instead lets the supervised loss on simulated mixtures regularize each update and anchor the target/non-target assignment of the DNN, while the real-recorded mixtures keep the model adapted to the target domain.
Because of memory limitations, we implement this by accumulating gradients over five mini-batches of size $1$, whose mixture types are drawn at random, so that each update typically aggregates both types of mixtures.
The parameters are updated once every five mini-batches.

\section{Experimental Setup}\label{exp_description}

We validate UltraM2M on the CHiME-$4$ dataset \cite{barker2016chime4,barker2015third}.
The corpus provides both simulated (SIMU) and real-recorded (REAL) mixtures for training and testing.
The signals are recorded with a tablet equipped with $6$ microphones, with the second microphone located on the rear and the others facing forward.
The recording environments include buses, cafeterias, streets, and pedestrian areas.
The training, validation, and test sets respectively contain $7{,}138$, $1{,}640$, and $1{,}320$ simulated mixtures, and $1{,}600$, $1{,}640$, and $1{,}320$ real mixtures. All signals are sampled at $16$ kHz.

\begin{figure}
    \centering
    \includegraphics[width=\linewidth]{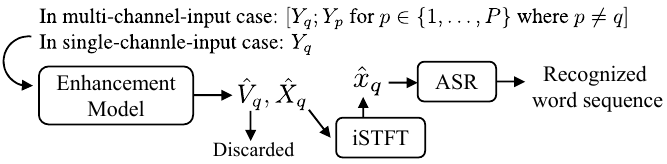}
    \caption{
    Robust ASR evaluation pipeline.
    }
    \label{fig:pipeline_enh_asr}
    \vspace{-0.2cm}
\end{figure}

For ASR evaluation, we reproduce a strong ASR system in ESPnet \cite{watanabe18_interspeech} released by Chang \textit{et al.} \cite{chang2022end}.
It is based on an encoder-decoder transformer trained on WavLM features \cite{chen2022wavlm} with a Transformer-based language model for decoding.
It achieves the best WER so far on unprocessed mixtures.
See Fig. \ref{fig:pipeline_enh_asr} for the entire pipeline for ASR evaluation, where the enhanced speech produced by our SE model is fed to the ASR model for recognition.

To train the UltraM2M model, we investigate three architecturally different ASR models: wav2vec 2.0 \cite{baevski2020wav2vec}, which is encoder-only and trained with connectionist temporal classification (CTC) loss; Whisper-Base \cite{radford2023robust}, which is encoder-decoder-based and trained with cross-entropy (CE) loss; and the same Chang \textit{et al.} model, which is encoder-decoder-based and is trained by combining CE and CTC losses.
Note that at run time, we always use Chang \textit{et al.}'s ASR model for decoding, while we can use a different ASR model for UltraM2M training.

For UltraM2M, the STFT window size is $32$ ms, the hop size is $8$ ms, and the square-root Hann window is used as the analysis window.
We employ TF-GridNet \cite{wang2023tf} as the DNN architecture of UltraM2M.
Using the symbols defined in Table I of \cite{wang2023tf}, we set $B=4, I=1, J=1, H=200, L=4, D=128$ and $E = 4$.
For the M2M algorithm, we set $I = 20, J=1$ and $\xi=10^{-2}$.
For the ASR backends, we set $\beta$ to $0,5$, $1.0$, and $0.05$ respectively for the Whisper-Base, wav2vec 2.0, and ESPnet backends, and we call these three variants \textbf{UltraM2M-Whis}, \textbf{UltraM2M-Wav2}, and \textbf{UltraM2M-ESPn}.
We train the DNN with $1$-, $2$-, and $6$-microphone inputs, while all $6$ microphones are used to compute the MC loss.
We take the fifth microphone as the reference microphone for the cases of using $1$- and $6$-channel inputs.
For the case of using $2$-channel input, we use the reference microphone and another microphone sampled from the other four front microphones.
DNSMOS OVRL \cite{reddy2022dnsmos} and word error rate (WER) are used as the evaluation metrics.
In training, we first train a DNN using the SuperM2M method \cite{wang2025superm2m} without using close-talk mixtures to initialize a common SE base model.
Starting from this initialization, we use UltraM2M to fine-tune the model.
The initial learning rates for SuperM2M and UltraM2M are respectively set to $10^{-3}$ and $10^{-4}$.
We employ early stopping with a patience of $2$ and a decay factor of $0.5$.
We also fine-tune the SE base model with the ASR loss alone to isolate the contribution of the MC loss.

\begin{table}[t]
\fontsize{6}{7}\selectfont
\centering
\captionsetup{justification=centering}
\sisetup{table-format=2.2,round-mode=places,round-precision=2,table-number-alignment = center,reset-text-series=false,text-series-to-math=true}
\caption{DNSMOS and WER results on CHiME-$4$.}
\vspace{-0.1cm}
\label{results_CHiME4_official}
\setlength{\tabcolsep}{3.5pt}
\begin{tabular}{
 l %
 l %
 S[table-format=1,round-precision=0] %
 S[table-format=1.2,round-precision=2] %
 S[table-format=1.2,round-precision=2] %
 S[table-format=1.2,round-precision=2] %
 S[table-format=2.2,round-precision=2] %
 S[table-format=1.2,round-precision=2] %
 S[table-format=2.2,round-precision=2] %
 S[table-format=2.2,round-precision=2] %
 S[table-format=2.2,round-precision=2] %
 }

\toprule

& & & \multicolumn{2}{c}{DNSMOS OVRL$\uparrow$} & \multicolumn{4}{c}{WER (\%)$\downarrow$} \\

\cmidrule(lr){4-5}\cmidrule(lr){6-9}

 & & {\multirow{2}{*}[-6pt]{\begin{tabular}[c]{@{}c@{}}Input\\\#mics\end{tabular}}} & {Val.} & {Test} & \multicolumn{2}{c}{Val.} & \multicolumn{2}{c}{Test} \\

\cmidrule(lr){4-4}\cmidrule(lr){5-5}\cmidrule(lr){6-7}\cmidrule(lr){8-9}

Row & System & & {REAL} & {REAL} & {SIMU} & {REAL} & {SIMU} & {REAL} \\

\midrule

0 & Mixture \cite{chang2022end} & 1 & 1.52 & 1.39 & 5.93 & 4.03 & 8.25 & 4.47  \\
[0.5ex]
\midrule
1 & IRIS \cite{chang2022end} & 1 & {-} & {-} & \bfseries 3.16 & 2.03 & \bfseries 6.12 & 3.92  \\ \hdashline\noalign{\vskip 0.5ex}
2a & Supervised & 1 & \bfseries 3.33 & \bfseries 3.18 & 3.53 & 2.15 & 8.05 & 4.83 \\
2b & SuperM2M \cite{wang2025superm2m} & 1 & 3.30 & 3.14 & 3.29 & 2.05 & 6.92 & 3.60 \\
2c & UltraM2M-Whis & 1 & 3.11 & 2.91 & 3.31 & \bfseries 1.86 & 6.94 & \bfseries 3.12 \\
2d & UltraM2M-Wav2 & 1 & 3.06 &  2.92 & 3.55 & 2.01 & 7.14 & 3.41 \\
2e & UltraM2M-ESPn & 1 & 2.95 & 2.71 & 3.34 & 1.91 & 6.77 & 3.13 \\
\midrule

3 & MultiIRIS~\cite{masuyama2023end} & 2 & {-} & {-} & 2.04 & 1.66 & \bfseries 2.04 & 2.65  \\
\hdashline\noalign{\vskip 0.5ex}
4a & Supervised & 2 &\bfseries  2.99 & \bfseries 2.83 & \bfseries 1.54 & 11.93 & 2.29 & 23.18 \\

4b & SuperM2M \cite{wang2025superm2m} & 2 & 1.81 & 1.64 & 1.57 & 2.71 & 2.22 & 2.85 \\
4c & UltraM2M-Whis & 2 & 2.71 & 2.53 & 1.90 & 1.55 & 2.12 & 2.21 \\
4d & UltraM2M-Wav2 & 2 &  2.84 & 2.62 & 1.95 & \bfseries 1.48 & 2.37 & \bfseries 2.16 \\
4e & UltraM2M-ESPn & 2 & 2.64 & 2.36 & 1.97 & 2.06 & 2.98 & 2.17 \\
4f & MIMO-Speech~\cite{Chang2019MIMOSPEECH} & 2 & 1.99 & 1.83 & 5.57 & 4.42 & 8.04 & 4.36 \\
\midrule

5 & MultiIRIS~\cite{masuyama2023end} & 6 & {-} & {-} & 1.22 & 1.33 & 1.24 & 1.77 \\
\hdashline\noalign{\vskip 0.5ex}
6a & Supervised & 6 & 2.36 & 2.03 & \bfseries 0.83 & 41.76 & 1.31 & 67.26 \\
6b & SuperM2M \cite{wang2025superm2m} & 6 & 1.84 & 1.63 & \bfseries 0.83 & 2.42 & 1.34 & 2.26 \\
6c & UltraM2M-Whis & 6 & \bfseries 2.66 & \bfseries 2.40 & 1.01 & \bfseries 1.41 & 1.19 & 1.90 \\
6d & UltraM2M-Wav2 & 6 & 2.65 & 2.39 & 1.03 & 1.51 & \bfseries 1.09 & 1.86 \\
6e & UltraM2M-ESPn & 6 & 2.40 & 2.08 & 1.25 & 1.67 & 1.50 & 1.92 \\
6f & MIMO-Speech~\cite{Chang2019MIMOSPEECH} & 6 & 1.54 & 1.43 & 4.25 & 6.93 & 4.68 & 8.24 \\

\bottomrule
\multicolumn{9}{l}{\textit{Notes}: \cite{wang2025superm2m} uses close-talk mixtures as an additional training loss.
Here close-talk mixtures are}\\
\multicolumn{9}{l}{assumed unavailable, so the SuperM2M results are reproduced without using them.}
\end{tabular}%
\vspace{2pt}
\footnotesize

\end{table}

\begin{table}[t]
\fontsize{6}{7}\selectfont
\captionsetup{justification=justified, singlelinecheck=false}
\caption{
Evaluation of MC loss
on CHiME-$4$ REAL sets with UltraM2M-Whis variants.
Rows $2$ and $6$ are respectively the $6$- and $1$-channel UltraM2M-Whis systems of Table \ref{results_CHiME4_official}.
}
\sisetup{round-mode=places, round-precision=2, group-digits=none, reset-text-series=false, text-series-to-math=true}
\label{tab:mc_generalization}
\setlength{\tabcolsep}{4pt}

\begin{tabular}{c l  S[table-format=1,round-precision=0]
  c S[table-format=2.2] S[table-format=2.2]
                S[table-format=1.2] S[table-format=2.2] }
\toprule
\multicolumn{2}{c}{} & {\multirow{2}{*}[-6pt]{\begin{tabular}[c]{@{}c@{}}\#input\\mics\end{tabular}}}& & \multicolumn{2}{c}{WER (\%)$\downarrow$} & \multicolumn{2}{c}{DNSMOS OVRL$\uparrow$}  \\
\cmidrule(lr){5-6}\cmidrule(lr){7-8}
 Row & System &   & Dataset & {Val.} & {Test} & {Val.} & {Test} \\
\midrule
1 & Supervised &   6 & SIMU & 41.76 & 67.26 & 2.36 & 2.03  \\
\hdashline\noalign{\vskip 0.5ex}
& UltraM2M-Whis \\
2 & \quad w/ MC &  6 & SIMU + REAL & \bfseries 1.41 & \bfseries 1.90 & \bfseries 2.66 & \bfseries 2.40  \\
3 & \quad w/o MC  & 6 & SIMU + REAL & 1.56 & 2.12 & 2.53 & 2.17 \\
4 & \quad w/ MC  & 6 & REAL & 1.59 & 2.32 & 2.60 & 2.29 \\
5 & \quad w/o MC  & 6 & REAL & 1.54 & 2.04 & 2.29 & 2.04 \\
\midrule
6 & \quad w/ MC & 1 & SIMU + REAL & \bfseries 1.86 & \bfseries 3.12 & \bfseries 3.11 & \bfseries 2.91 \\
7 & \quad w/o MC & 1 & SIMU + REAL & 2.53 & 4.76 & 2.46 & 2.23 \\
\bottomrule
\end{tabular}%
\end{table}
\begin{figure}
    \centering
    \includegraphics[width=0.8\linewidth]{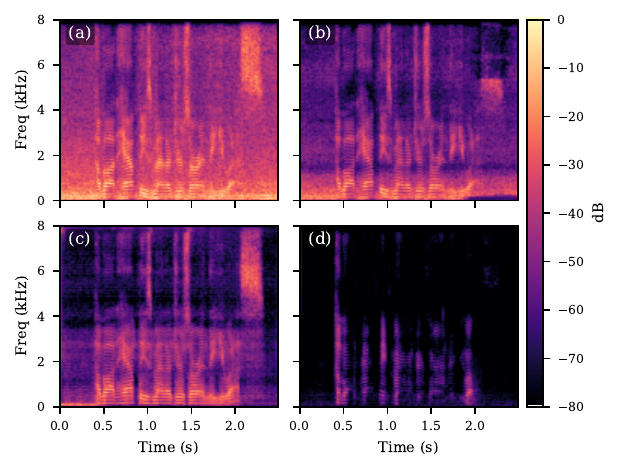}
    \vspace{-0.4cm}
    \caption{
    Enhancement results of an example CHiME-$4$ mixture: (a) Unprocessed mixture; (b) SuperM2M; (c) UltraM2M-Whis and (d) Supervised.}
    \vspace{-0.4cm}
    \label{fig:spectrograms}
\end{figure}

\section{Evaluation Results and Discussion}\label{eval_results_description}
Table \ref{results_CHiME4_official} reports the DNSMOS-OVRL and WER results on the CHiME-$4$ simulated and real-recorded validation and test sets.
The purely supervised model (denoted as ``Supervised'') is trained on simulated mixtures only.
We observe that it generalizes poorly to real-recorded data in terms of ASR performance, consistent with the observations in \cite{wang2025superm2m}, whereas SuperM2M clearly improves on real-recorded data, showing the effectiveness of M2M training.
Relative to SuperM2M, UltraM2M reduces WER on the REAL test set for every ASR backend and channel count, and in conditions with $2$- and $6$-channel input it also improves DNSMOS.
This simultaneous gain in perceptual quality and recognition accuracy confirms the effectiveness of UltraM2M. 

Fig. \ref{fig:spectrograms} provides a mixture example to qualitatively illustrate the effectiveness of UltraM2M.
The mixture spectrogram is dominated by background noise.
SuperM2M suppresses the noise to some extent, but the speech harmonics remain unclear.
UltraM2M, driven by an ASR loss, suppresses the noise more aggressively and yields clearer harmonic structure, while retaining more of the original signal structure than the purely supervised model, which tends to over-suppress especially in mismatched test conditions.
Note that, although the three ASR backends differ in architecture and in the training criterion they provide, the Whisper-Base and wav2vec 2.0 variants deliver consistently better perceptual quality and comparable or better recognition accuracy on the real-recorded sets, indicating that the proposed training does not depend on a specific ASR backend.

MIMO-Speech \cite{Chang2019MIMOSPEECH} is another weakly-supervised speech enhancement method, also leveraging text transcripts as supervision.
However, its enhancement result is obtained via time-invariant beamforming which often has limited performance in conditions with strong reverberation, diffuse noises and limited microphone counts.
We reproduce it on CHiME-$4$ using TF-GridNet as the masking network for its MVDR beamformer, the same wav2vec 2.0 backend as UltraM2M-Wav2, and the same combination of simulated and real-recorded training mixtures.
Unlike MIMO-Speech, the MC loss is designed for noisy-reverberant environments and can work reasonably well even for input signals with a limited number of microphones.
UltraM2M obtains consistently better results on both metrics, showing more robust effectiveness.
Moreover, UltraM2M outperforms IRIS \cite{chang2022end} and MultiIRIS \cite{masuyama2023end} in $1$- and $2$-channel conditions, and remains competitive with MultiIRIS in $6$-channel conditions, where IRIS and MultiIRIS are the previous best.

One further observation from Table \ref{results_CHiME4_official} is that the ASR loss helps more in the multi-channel settings, where SuperM2M's perceptual quality is markedly lower than in the monaural setting.
With more input channels but only a weak MC constraint, the DNN receives less guidance on which spectro-temporal structure to retain and which to discard, and the ASR loss resolves much of this ambiguity, thereby preventing the perceptual quality from collapsing.
Overall, the transcript-level supervision improves recognition consistently and yields additional noise reduction in the multi-channel settings, while the MC loss and the supervised branch are designed to prevent this transcript-driven training from drifting into a degraded regime.

Table \ref{tab:mc_generalization} isolates the contribution of each constraint.
Adding the MC loss improves DNSMOS OVRL consistently, both when real-recorded mixtures are used alone and when simulated mixtures are added.
Its effect on WER, however, depends on the data regime: with real-recorded mixtures only, the MC loss trades a small amount of recognition accuracy for that perceptual gain.
When simulated mixtures are also used, in contrast, the MC loss is what makes the system generalize; without it, the gap between the validation and the test results widens.
This motivates keeping supervised training on simulated mixtures as an auxiliary constraint, rather than training on real-recorded mixtures alone.

\section{Conclusions}\label{conclusion_description}

We have proposed UltraM2M, a weakly-supervised SE method that leverages the text transcripts of real-recorded mixtures through pretrained, frozen ASR models.
M2M training and the co-learning strategy with simulated mixtures provide the constraints that make this weak supervision effective.
We find that text transcripts are helpful for weakly-supervised SE, and that the enhancement quality improves when they are used under suitable constraints.
UltraM2M could therefore be a practical solution for realistic SE problems where clean reference speech is usually unavailable, but text transcripts are cheap to annotate and readily accessible.


{\footnotesize
\bibliographystyle{IEEEtran}
\bibliography{references.bib}
}

\end{document}